\documentclass[twocolumn,amsmath,ammsymb]{revtex4-1}
\usepackage{epsfig}
\usepackage{amsmath}
\usepackage{amssymb}
\usepackage{graphicx}
\usepackage{dcolumn}
\usepackage{soul}
\usepackage{bm}
\usepackage[usenames]{color}
\usepackage[breaklinks]{hyperref}
\hypersetup{colorlinks=true,allcolors=blue}

\begin{document}

\preprint{}
 
\title{Phase diagram and orbital Chern insulator in twisted double bilayer graphene}
\author{Yi-Xiang Wang$^{1,2}$} \email{wangyixiang@jiangnan.edu.cn}
\author{Fuxiang Li$^2$} \email{fuxiangli@hnu.edu.cn}
\author{Zi-Yue Zhang$^1$}
\affiliation{$^1$School of Science, Jiangnan University, Wuxi 214122, China.}
\affiliation{$^2$School of Physics and Electronics, Hunan University, Changsha 410082, China}
\date{\today}

\begin{abstract}
Compared with twisted bilayer graphene, twisted double bilayer graphene (TDBG) provides another important platform to realize the moir\'e flat bands.  In this paper, we first calculate the valley Chern number phase diagram of TDBG in the parameter space spanned by the twist angle and the interlayer electric potential.  To include the effects of interactions, we then phenomenologically introduce the spin-splitting and valley-splitting.  We find that when the valley splitting is larger than the bandwidth of the first conduction band so that a gap is opened and the spin splitting is relatively weak, the orbital Chern insulator emerges at half-filling, associated with a large orbital magnetization (OM).  Further calculations suggest that there is no sign reversal of the OM when the Fermi energy goes from the bottom to the top of the half-filling gap, as the OM remains negative in both AB-AB stacking and AB-BA stacking.  The implications of our results for the ongoing experiments are also discussed.
\end{abstract}

\maketitle

\section{Introduction}
 
The recent discovery of the correlated insulator states~\cite{Y.Cao2018a}, superconductivity~\cite{Y.Cao2018b, M.Yankowitz} as well as quantum anomalous Hall (QAH) state~\cite{A.L.Sharpe, M.Serlin} in twisted bilayer graphene (TBG) have drawn significant attentions.  In TBG, the spatial variation of interlayer coupling modifies the Dirac linear band structure of graphene in such a way that the band dispersion is almost completely suppressed at the so-called magic angle~\cite{R.Bistritzer}.  As the bandwidth $w$ of the flat band is sufficiently narrow, it is possible to achieve the situation $\frac{U}{w}\gg1$ so that the effective Coulomb interaction $U$ dominates the system.  The interaction provides the possible mechanism for the observed correlated insulator states and superconducting states upon charge doping~\cite{B.Lian, F.Wu2018, B.Roy}.  It was further revealed that the low-energy flat bands could have well-defined valley Chern numbers, which can host a number of fascinating many-body phenomena, including the fractional QAH effects~\cite{Y.H.Zhang} .  
 
This novel twist-angle degree of freedom and its control could be generalized to other two-dimensional system, where similar correlated physics may also be exhibited.  It has been demonstrated that twisted double bilayer graphene (TDBG)~\cite{G.W.Burg, C.Shen, Y.Cao2020, X.Liu, M.He} and ABC-stacked trilayer graphene on hexagonal boron nitride (hBN) supperlattices~\cite{G.Chen2019a,G.Chen2019b} can provide another important moir\'e systems with strong correlation effect.  TDBG refers to a pair of bilayer graphene twisted with each other by a small angle $\theta$.  There may exist two different stacking types for TDBG, AB-AB stacking and AB-BA stacking, both of which will be considered in this paper.  Unlike TBG, the isolated flat moir\'e band in TDBG can appear when an out-of-plane electric field is applied on the system~\cite{M.Koshino2019}.  More importantly, as the bilayer graphene becomes gapped under the  electric field, the opposite Berry curvatures at the two valleys can be accumulated~\cite{I.Martin, F.Zhang}, leading to the change of the band Chern number.  As both the twist angle and the electric field can be well controlled in experiment, the study of the driven Chern number phase diagram in TDBG is meaningful and gives the first motivation of the present work. 

In TBG, the topological flat moir\'e bands are closely connected to the large orbital magnetizations (OMs), which may give rise to an orbital Chern insulator (OCI) state once the valley symmetry is broken.  In fact, the OCI has been successfully observed at $n=\frac{3}{4}n_s$ filling of TBG when aligned with the hBN substrate around the magic angle~\cite{A.L.Sharpe, M.Serlin}, with $n_s=\frac{4}{S_M}$ being the density corresponding to fully filling one moir\'e band, the factor $4$ accounting for the spin and valley flavors and $S_M$ denoting the size of the unit moir\'e cell.  The time-reversal symmetry (TRS) breaking mechanism of OCI in TBG can be attributed to the condensation of the electrons in the momentum space, where the many-body interaction drives the spontaneous valley polarization.  Another important system called the spin Chern insulator was observed in Cr-doped (Bi,Sb)$_2$Te$_3$ thin film~\cite{C.Z.Chang}, where the TRS is broken by the local spin moments that are ordered ferromagnetically due to the exchange interaction.  Both the orbital and spin Chern insulators are quite different from the Chern insulator identified in the original Haldane model~\cite{Haldane}, where the TRS is broken by the local staggered magnetic flux in a unit cell, leading to the Berry curvatures of the same sign around the two valleys.  

In this paper, we will study under what conditions the OCI can be realized in TDBG.  Although in a unit cell, TDBG has two times the sublattices than TBG, the band counting is the same and each moir\'e band accommodates four electrons per unit moir\'e cell when the spin-degeneracy and valley-degeneracy are preserved.  If the number of electrons per moir\'e band is a multiple of four, gaps may appear.  However, when the Fermi energy lies in the gap, the Chern number vanishes due to the opposite contributions from the two valleys that are connected by the TRS.  Thus to observe the nontrivial Chern insulator phase in TDBG, the four-fold degeneracy of the moir\'e band needs to be broken, which may be achieved by the spin-splittings and valley-splittings.  Moreover, the magnetic property studied in TBG revealed that the OM may reverse its sign when doping the system from the bottom to the top of the insulating gap, and suggested that this is quite common in the OCI~\cite{J.Zhu}.  The external perpendicular magnetic field $\boldsymbol B$ favors the state with magnetization $\boldsymbol M$ aligned in the same direction, leading to the stronger resistive signal in the transport experiment.  Thus the sign reversal of the OM can drive a reversal of the valley polarization when the Fermi energy crosses the gap, enabling the electrical switching of a magnetic state in TBG in when a fixed magnetic field is present~\cite{H.Polshyn}. Then a natural question arises that what is the magnetic property in TDBG when the OCI is realized, which will also be explored in this paper.  

Our main findings are as follows: (i) By using the Fukui's algorithm, we perform accurate calculations of the Chern numbers of the first valence and first conduction band, $(C_{v1},C_{c1})$, and obtain a reliable phase diagram of TDBG as a function of the twist angle and the electric potential, especially for the regions where the neighboring moir\'e bands are overlapped.  We find that the phase diagrams for AB-AB stacking and AB-BA stacking show significant discrepancies.  Since the $(C_{v1},C_{c1})=(2,-2)$ phase in the AB-BA stacking and the $(1,1)$ phase in the AB-BA stacking share similar broad parameter regions, and in the two phases the bands are relatively flat, our studies are mainly focused on these two phases.  (ii) By phenomenologically introducing the interaction-induced valley splitting and spin splitting, the OCI state is found at half-filling $n=\frac{1}{2}n_s$ of the first conduction band and is associated with the large OM.  The condition is that, the valley splitting should be larger than the bandwidth so as to open a gap at half-filling, and the spin splitting should be relatively weak.  (iii) For the OCI in TDBG, our calculations suggest that there is no sign reversal of the OM when the Fermi energy goes from the bottom to the top of the half-filling gap, as the OM remains negative in both AB-AB stacking and AB-BA stacking.  Our study could help explore the twist-angle and electric-field modulated topological phases of matter in the flat-band twisted superlattice systems.

\section{Twisted double bilayer graphene model}

\begin{figure}
	\includegraphics[width=8.8cm]{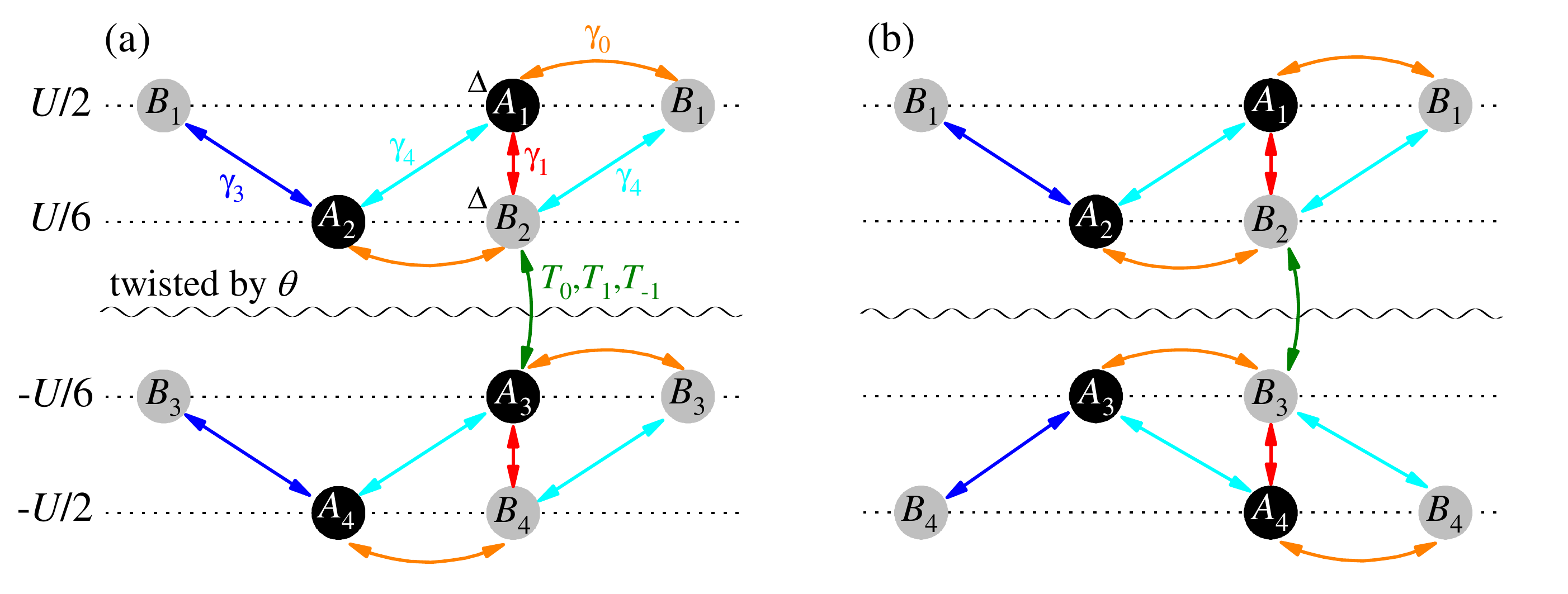}
	\caption{(Color online) Schematics of TDBG with AB-AB stacking (a) and AB-BA stacking (b).  The hopping integrals between the neighboring sublattices, $\gamma_{0,1,3,4}$, are shown with arrows in different colors.  For the dimer sublattices, there is a small on-site energy $\Delta$.  The layer-dependent potentials are also labeled.}
	\label{Fig1}
\end{figure}
 
Bilayer graphene is composed of a pair of monolayer graphene, where four sublattices are included in a unit cell, labeled as $A_1$, $B_1$ in the upper layer and $A_2,B_2$ in the bottom layer.  In bilayer graphene, the most stable configuration is AB or BA stacking, which is also the stacking structure of 3D bulk graphite~\cite{Neto}.  In AB (BA) stacking, the $A_1$ $(B_1)$ sublattice of the upper layer is located on the top of the $B_2$ $(A_2)$ sublattice of the lower layer, leading to a small on-site energy $\Delta$ for these dimer sublattices.  The other two sublattices, $B_1$ and $A_2$ ($A_1$ and $B_2$) are directly above or below the hexagon center of the other layer.  The schematics of TDBG are shown in Fig.~\ref{Fig1}, with the two stacking types, AB-AB stacking and AB-BA stacking. 
 
To describe the single-particle band structure of TDBG, we adopt the commonly used tight-binding model~\cite{R.Bistritzer, M.Koshino2018, M.Koshino2019, J.Y.Lee, J.Liu, N.R.Chebrolu}.  In the eight-component basis  $(c_{A_1},c_{B_1},c_{A_2},c_{B_2},c_{A_3},c_{B_3},c_{A_4},c_{B_4})^T$, the Hamiltonian at small twist angle $\theta$ is
\begin{align}
H_{AB-AB}(\boldsymbol k)=\begin{pmatrix}
h_0(\boldsymbol k_1)& g^\dagger(\boldsymbol k_1)&&
\\
g(\boldsymbol k_1)& h_0'(\boldsymbol k_1)& T^\dagger&
\\
& T& h_0(\boldsymbol k_2)& g^\dagger(\boldsymbol k_2)
\\
&& g(\boldsymbol k_2)& h_0'(\boldsymbol k_2)
\end{pmatrix}+V, 
\end{align}
and 
\begin{align}
H_{AB-BA}(\boldsymbol k)=\begin{pmatrix}
h_0(\boldsymbol k_1)& g^\dagger(\boldsymbol k_1)&&
\\
g(\boldsymbol k_1)& h_0'(\boldsymbol k_1)& T^\dagger&
\\
& T& h_0'(\boldsymbol k_2)& g(\boldsymbol k_2)
\\
&& g^\dagger(\boldsymbol k_2)& h_0(\boldsymbol k_2) 
\end{pmatrix}+V. 
\end{align} 
Here $\boldsymbol k_l=R(\pm\frac{\theta}{2})(\boldsymbol k-\boldsymbol K_\xi^l)$ is the in-plane momentum, with $R(\theta)$ being the two-dimensional rotation matrix and the sign $\pm$ for the top ($l=1$) and bottom $(l=2)$ bilayer graphene, respectively.  $\xi=\pm1$ is the valley index and $\boldsymbol K_\xi^l$ is the corresponding Dirac point.  Both $h_0(\boldsymbol k)$ and $h_0'(\boldsymbol k)$ describe the intralayer hoppings between sublattices $A$ and $B$, while $g(\boldsymbol k)$ denotes the coupling between the two layers in bilayer graphene.  These $2\times2$ submatrices are written as
\begin{align}
&h_0(\boldsymbol k)=\begin{pmatrix}
\Delta& -\gamma_0f(\boldsymbol k)
\\
-\gamma_0f^*(\boldsymbol k)& 0
\end{pmatrix}, 
\\
&h_0'(\boldsymbol k)=\begin{pmatrix}
0& -\gamma_0f(\boldsymbol k)
\\
-\gamma_0f^*(\boldsymbol k)& \Delta
\end{pmatrix}, 
\\
&g(\boldsymbol k)=\begin{pmatrix}
\gamma_4f(\boldsymbol k)& \gamma_3f^*(\boldsymbol k)
\\
\gamma_1& \gamma_4f(\boldsymbol k)
\end{pmatrix}, 
\end{align} 
where $\gamma_0$ is the nearest-neighbor hopping integral and $f(\boldsymbol k)=\sum_i e^{-i\boldsymbol k\cdot \boldsymbol \delta_i}$, with $\boldsymbol\delta_1=a_0(0,-\frac{1}{\sqrt3})$, 
$\boldsymbol\delta_2=a_0(-\frac{1}{2},\frac{1}{2\sqrt3})$,
$\boldsymbol\delta_3=a_0(\frac{1}{2},\frac{1}{2\sqrt3})$ denoting the vectors pointing from sublattice $A$ to $B$, and $a_0$ being the lattice constant.  We can expand $f(\boldsymbol k)$ around the Dirac points $\boldsymbol K_\pm=(\pm\frac{4\pi}{3a_0},0)$ as $f(\boldsymbol K_\pm+\boldsymbol k)=\frac{\sqrt3a_0}{2}(\mp k_x+ik_y)$.  In $g(\boldsymbol k)$, the parameter $\gamma_3$ represents the trigonal warping of the energy bands and $\gamma_4$ accounts for the electron-hole asymmetry in bilayer graphene~\cite{E.McCann}.  The tight-binding parameters are labeled in detail in Fig.~\ref{Fig1}.   We use the parameters that are extracted from the \textit{ab initio} results of Ref.~\cite{J.Jung2014}, $\gamma_0=2610$ meV, $\gamma_1=361$ meV, $\gamma_3=283$ meV, $\gamma_4=138$ meV and $\Delta=15$ meV. 

The term $V$ in the Hamiltonian describes the effect of the out-of-plane perpendicular electric field, as it can induce the interlayer asymmetric electric potential.  In bilayer graphene, the electric potential difference between the two layers can open a gap in the parabolic touching bands~\cite{E.McCann}.  We assume that the electric potential drop between the neighboring layers is uniform, $U_i-U_{i+1}=\frac{U}{3}$, as shown in Fig.~\ref{Fig1}.  Specifically,
\begin{align}
V=\begin{pmatrix}
\frac{U}{2} I&  
\\
& \frac{U}{6} I&  
\\
&& -\frac{U}{6} I&  
\\
&&& -\frac{U}{2} I
\end{pmatrix}, 
\end{align}
with $I$ being the $2\times2$ unit matrix.  In experiment, the electric potential in TDBG can be effectively tuned by the top and back gates~\cite{Y.Cao2020, C.Shen, X.Liu, G.W.Burg}. 

The tunneling $T(\boldsymbol r)$ between the top and bottom bilayer graphene varies with the moir\'e period and is written as~\cite{R.Bistritzer, M.Koshino2018}
\begin{align} 
&T(\boldsymbol r)=T_0+e^{-i\boldsymbol b_+\cdot\boldsymbol r}T_{+1}
+e^{-i\boldsymbol b_-\cdot\boldsymbol r}T_{-1}, 
\\
&T_j=w_0\sigma_0+w_1\text{cos}(j\frac{2\pi}{3})\sigma_x
+w_1\text{sin}(j\frac{2\pi}{3})\sigma_y, 
\end{align}
where $\boldsymbol b_\pm=\frac{4\pi}{\sqrt3a_M}(\pm\frac{1}{2},\frac{\sqrt3}{2})$ are the moir\'e reciprocal lattice vectors and $a_M=\frac{a_0}{2\text{sin}\frac{\theta}{2}}$ denotes the moir\'e period.  Because the moir\'e period is much larger than the lattice constant, $a_M\gg a_0$, the intervalley scatterings can be safely ignored and we treat the two valleys separately.  Moreover, as the two valleys are connected by the TRS, we mainly focus on $\boldsymbol K$ valley, while the physics of $\boldsymbol K'$ valley can be obtained by the TR operation.  $w_0$ and $w_1$ are the two tunneling parameters, which in general are unequal due to the layer corrugation in the moir\'e pattern.  We take $w_0=79.5$ meV and $w_1=97.5$ meV~\cite{M.Koshino2018} in the following calculations.  

The moir\'e potential reconstructs the original Dirac linear bands into the small moir\'e Brillouin zone (MBZ).  Numerically, the band structures can be effectively calculated by using the plane-wave expansions~\cite{R.Bistritzer}.  For each momentum $\boldsymbol k$, we use the basis that include the states of $(2M+1)\times(2M+1)$ momentum points: $\boldsymbol k+n_1\boldsymbol b_+ + n_2(\boldsymbol b_+-\boldsymbol b_-)$, where $-M\leq n_1,n_2\leq M$ are integers.  In the calculations, we choose $M=4$ to achieve results that are well convergent.

\section{Phase diagram} 

\begin{figure*}
	\centering
	\includegraphics[width=18cm]{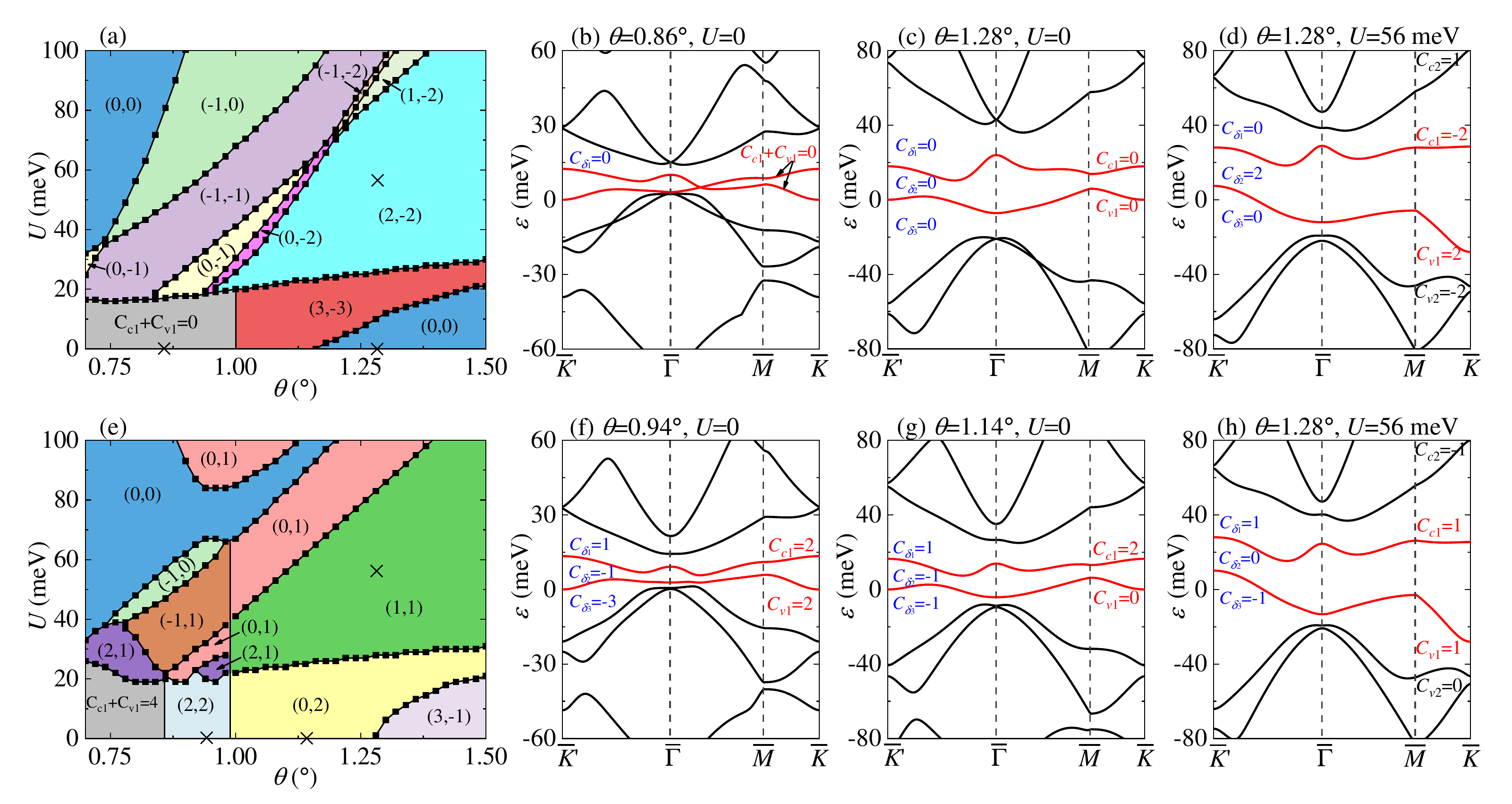}
	\caption{(Color online) The $\boldsymbol K$ valley Chern number phase diagram of TDBG in the parameter space $(\theta,U)$ with AB-AB stacking (a) and AB-BA stacking (e).  The different phases are characterized by the Chern numbers of the first valence and conduction band $(C_{v1},C_{c1})$.  (b)-(d) and (f)-(h) are the typical moir\'e bands along the high-symmetry line, $\bar{K}'\rightarrow{\bar\Gamma}\rightarrow{\bar M}\rightarrow{\bar K}$, in the MBZ, with the parameter points being marked by the crosses in (a) and (e), respectively.  Note that Chern numbers are labeled in each figure. }
	\label{Fig2}
\end{figure*}

First we calculate the $\boldsymbol K$ valley Chern number phase diagram of TDBG, as the moir\'e bands are generically topological and can carry nonzero Chern number~\cite{Y.H.Zhang}.  For the $n$th band, its Chern number is defined as an integration over the MBZ~\cite{D.Xiao} 
\begin{align} 
C_n=\frac{i}{2\pi}\int_{\text{MBZ}}
d^2\boldsymbol k \langle \frac{\partial u_{n\boldsymbol k}}{\partial \boldsymbol k} |\times
|\frac{\partial u_{n\boldsymbol k}}{\partial \boldsymbol k}\rangle, 
\end{align}
with $|u_{n\boldsymbol k}\rangle$ being the Bloch wavefunction.  The Chern number can be numerically calculated by using the Fukui's algorithm, in which the BZ is divided into many disconnected sectors and a unique topological invariant is assigned to each sector.  Then the Chern number is written as~\cite{T.Fukui, Y.X.Wang}
\begin{align}
C_n=&\frac{1}{2\pi}\sum_i \text{Im}\Big[ \text{ln}
\Big(\langle u_{n\boldsymbol k_i}^1|u_{n\boldsymbol k_i}^2\rangle
\langle u_{n\boldsymbol k_i}^2|u_{n\boldsymbol k_i}^3\rangle
\nonumber\\
&\times
\langle u_{n\boldsymbol k_i}^3|u_{n\boldsymbol k_i}^4\rangle
\langle u_{n\boldsymbol k_i}^4|u_{n\boldsymbol k_i}^1\rangle
\Big)
\Big], 
\end{align}
where the summation is to be taken over all disconnected sectors, and $|u_{n\boldsymbol k_i}^j\rangle$ ($j=1,2,3,4$ in anticlockwise direction) is the $n$th wavevector corresponding to the four vertices in the $i$th sector.  The advantage of the Fukui's algorithm is that it can calculate the Chern number of a specific band in a reliable way, even when neighboring bands are overlapped, as long as the bands do not touch with each other.  We label the Chern number of the $n$th valence (conduction) bands as $C_{vn}$ ($C_{cn}$).  The first valence and first conduction bands will be focused on and the Chern numbers $(C_{v1},C_{c1})$ are used to distinguish the different phases, as they can undergo multiple changes at the high-symmetry points in the MBZ. 

When the electric potential $U$ in TDBG reverses its direction, we find that for AB-AB stacking, the Chern number of the $n$th band turns to its opposite value, $C_n(-U)=-C_n(U)$, while for AB-BA stacking, it will not change, $C_n(-U)=C_n(U)$.  This is because in AB-AB stacking, the $C_{2x}$ symmetry is broken by the electric potential, while in AB-BA stacking, the $C_{2x}$ symmetry is maintained.  Specifically, if we rotate the TDBG system with negative $U$ by 180$^\circ$ along the $x-$axis in the 2D plane, for AB-AB stacking, the rotated system becomes BA-BA stacking with positive $U$.  As the chirality of the massive bands changes, it makes the Chern number reverse to its opposite value.  However, for AB-BA stacking, the rotated system returns to its origin with positive $U$ and thus the Chern number remains unchanged.  This property may be used in experiment to judge whether the chiralities of the two stacked bilayer graphene are the same or not. 

The $\boldsymbol K$-valley Chern number phase diagram of TDBG is plotted in Fig.~\ref{Fig2} with AB-AB stacking in (a) and AB-BA stacking in (e), where the different phases are labeled in different colors.  We can see that the Chern number are tunable up to $\pm3$.  Clearly, there are significant discrepancies of the two phase diagrams in the two stacking types.  The typical moir\'e band structures along the high-symmetry line in the MBZ are plotted in Figs.~\ref{Fig2}(b)-(d) and (f)-(g).  We define the bandgap between the first conduction and second conduction band as $\delta_1$, the bandgap between the first conduction and valence band as $\delta_2$ and the bandgap between the first valence and second valence band as $\delta_3$.  When $\theta$ is small and $U$ is lower than 20 meV, the first conduction and valence band touch with each other [e.g., see Fig.~\ref{Fig2}(b)], making $\delta_1$ unopened and the Chern number ill-defined.  However, the bandgaps $\delta_1$ and $\delta_3$ are opened and can protect the sum of the Chern numbers.  So we use $C_{c1}+C_{v1}$ to characterize these phases.  As shown in Figs.~\ref{Fig2}(a) and (e), the phase of $C_{c1}+C_{v1}=0$ and $C_{c1}+C_{v1}=4$ spans the lower left region of the phase diagram, respectively. 

For AB-AB stacking, when $U=0$ and $\theta>1^\circ$, the two touching bands are separated.  The increasing $\theta$ drives the system first enter the $(3,-3)$ phase and then the $(0,0)$ phase [Fig.~\ref{Fig2}(c)].  As the two lowest bands are separated from the higher bands, the summation of the Chern numbers remains zero, $C_{c1}+C_{v1}=0$.  We can see that the Chern number in the middle bandgap, $C_{\delta_2}$, which is defined as the summation of the band Chern number below the bandgap $\delta_2$, is also zero.  The increasing $U$ can drive the bands touch at the high-symmetry points and then separate, resulting in the change of $C_{c1}$ or $C_{v1}$.  Note that the $(2,-2)$ phase [Fig.~\ref{Fig2}(d)] spans a broad parameter region in the phase diagram, meaning that it remains unchanged to the small variations of $U$ and $\theta$.  Moreover, the $(2,-2)$ phase represents a valley Chern insulator as $C_{\delta_2}=2$. 

For AB-BA stacking, when $U=0$ and $0.86^\circ<\theta<0.98^\circ$, the Chern numbers become $(2,2)$, but the direct gap is too small or even does not exist [Fig.~\ref{Fig2}(f)].  When $U<66$ meV and $\theta\sim 0.98^\circ$, the first valence band will touch with the higher valence band at the $\bar\Gamma$ point.  Thus the phase transitions happen and a vertical phase boundary at $\theta\sim0.98^\circ$ is seen, where $C_{v1}$ varies but $C_{c1}$ keeps unchanged.  For example, at $U=0$, the increasing $\theta$ drives the $(2,2)$ phase enter the $(0,2)$ phase [Fig.~\ref{Fig2}(g)].  It shows that $C_{\delta_2}=-1$ and is distinct from AB-AB stacking.  We also note that the $(1,1)$ [Fig.~\ref{Fig2}(h)] phase behaves as a trivial insulator as $C_{\delta_2}=0$.  More importantly, it spans a similar broad parameter region in the phase diagram as the $(2,-2)$ phase in AB-AB stacking.

In a previous work~\cite{N.R.Chebrolu}, the valley Chern number phase diagrams of TDBG are obtained from the TKNN formula [see Eq.~(11) below], which are partly agreement with our results.  The differences between them mainly lie in the parameter regions where the neighboring bands related to the first conduction and first valence bands are overlapped [see Appendix A].  Our results show that when the neighboring bands are overlapped, the different computational methods may lead to different results [see Appendix B].  As is known, when there is a direct bandgap between the neighboring topological bands and the Fermi energy lies in it, the TKNN formula [see Eq.~(12) below] can express the anomalous Hall conductivity (AHC) $\sigma_H$ (in unit of  $\frac{e^2}{h}$) as a quantized value, which equals to the Chern number in the gap $C_\delta$.  Then the band Chern number is determined and equals to the Chern number in the above bandgap minus the Chern number in the below bandgap.  In this case, we have checked that the Chern number results obtained by using the Fukui's algorithm and the TKNN formula are consistent with each other.  When the the bands are overlapped, the quantized $\sigma_H$ will not appear.  In this case, the Chern number judgement from the TKNN formula may be inconvenient. However, we suggest that the Fukui's algorithm is still valid for determining the Chern number, as long as the neighboring bands do not touch with each other.  The valley Chern number phase diagram was also reported in another work~\cite{J.Y.Lee}, but was only about the first conduction band $C_{c_1}$.  It is worthy pointing out that our phase diagrams are also consistent with two recent studies~\cite{J.Liu, F.Wu2020}, where the Chern numbers in TDBG are presented for some specific parameter points of $(\theta,U)$.

\begin{figure}
	\includegraphics[width=8.4cm]{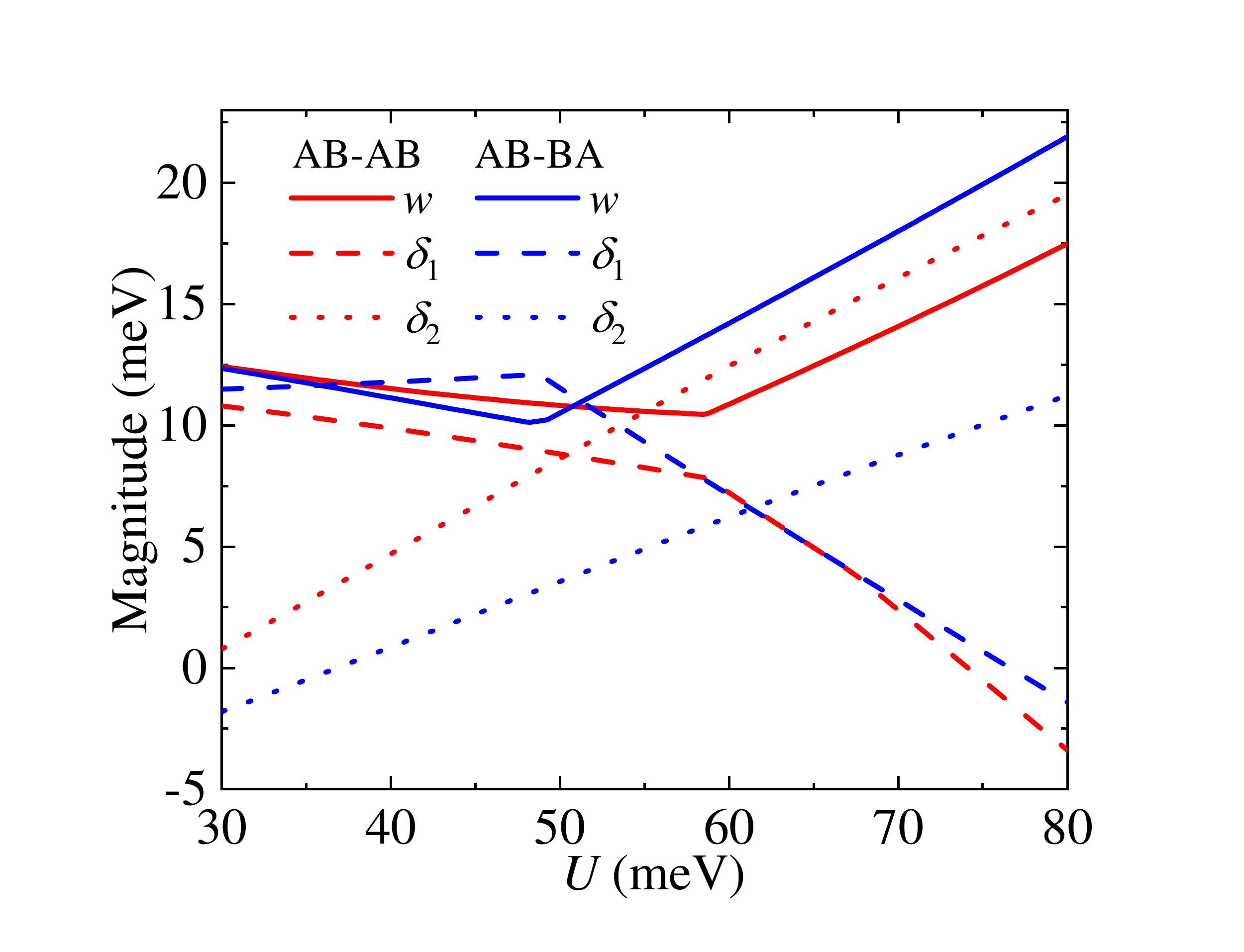}
	\caption{(Color online) The flatness $w$ of the first conduction band, the gaps $\delta_1$ and $\delta_2$ vs the interlayer potential $U$ in TDBG, where both AB-AB stacking and AB-BA stacking are considered.  We set the twist angle $\theta=1.28^\circ$.}
	\label{Fig3}
\end{figure}

We further study the evolution of the first conduction band with the electric potential $U$, as it can be well isolated from other bands.  In Fig.~\ref{Fig3} with the fixed $\theta=1.28^\circ$, we plot the flatness of the first conduction band $w$, the bandgaps $\delta_1$ and $\delta_2$ as functions of $U$.  It shows that for both AB-AB and AB-BA stacking types, these quantities exhibit similar trends.  Around $U=30$ meV, we have $w=12.5$ meV, which is comparable to $\delta_1$, whereas $\delta_2$ is close to zero.  At large $U$, both $w$ and $\delta_2$ increase while $\delta_1$ decreases, meaning that the first conduction band becomes wider and moves closer to the second conduction band.  In the extremal case when $U$ is sufficiently high (low), the neighboring bands are overlapped and $\delta_1$ ($\delta_2$) becomes negative.  These results can be used to explain the recent resistance measurements in TDBG with $\theta$ being around $1.3^\circ$~\cite{Y.Cao2020, X.Liu, M.He}, where the insulating state at charge neutrality (corresponding to $\delta_2$) strengthens with the electric potential $U$, while the $n=+n_s$ insulating state (corresponding to $\delta_1$) is weakened and eventually disappears with the increasing $U$.  In addition, the observed asymmetric change of insulating states at $n=+n_s$ and $n=-n_s$ versus $U$~\cite{Y.Cao2020, X.Liu, M.He} can be attributed to the broken electron-hole symmetry in the TDBG moir\'e bands. 

\begin{figure*}
	\centering
	\includegraphics[width=18.2cm]{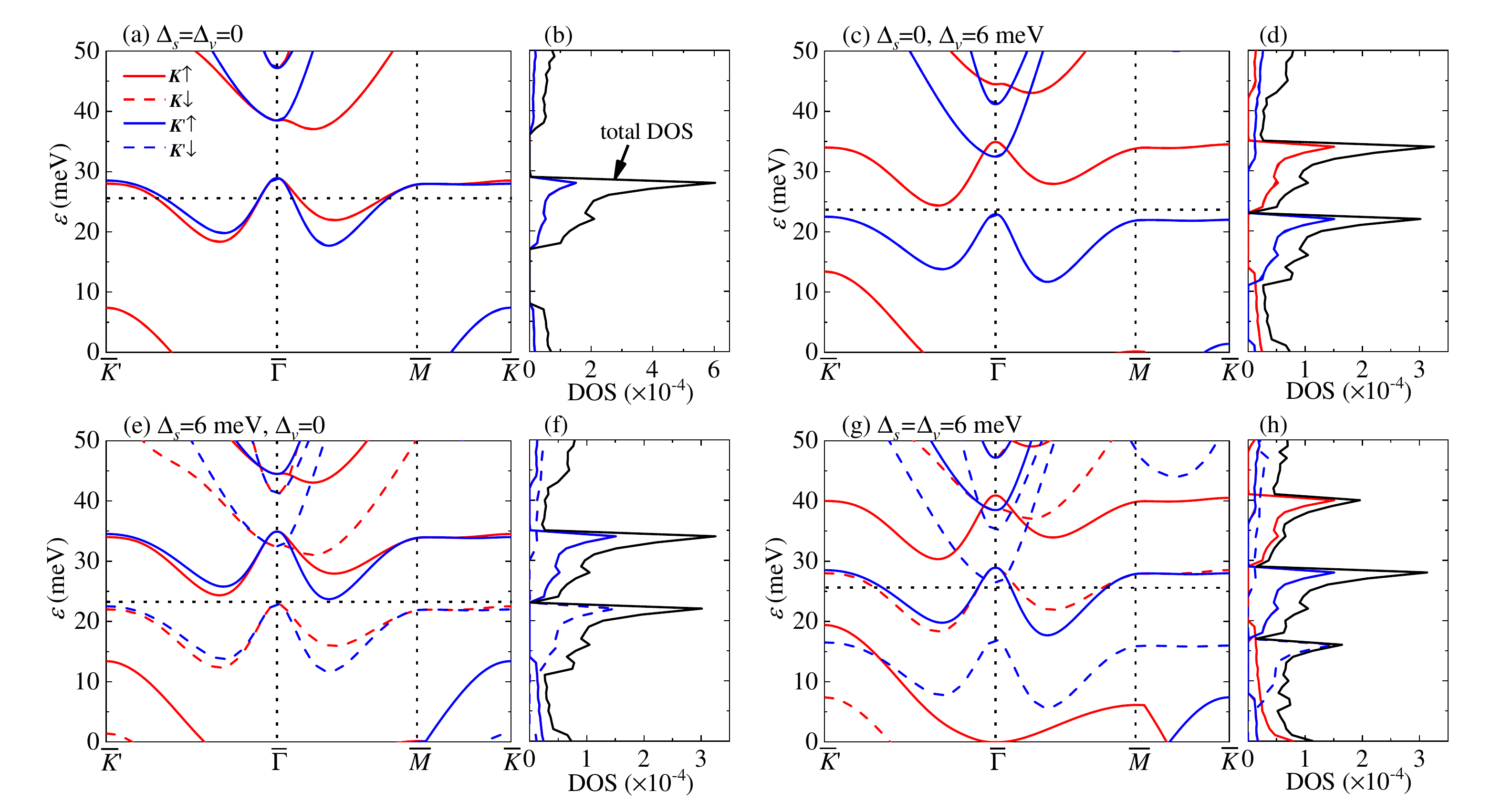}
	\caption{(Color online) The moir\'e bands (a), (c), (e) and (g), and the DOS (b), (d), (f), and (h) of TDBG with AB-AB stacking, with the different spin splitting $\Delta_s$ and valley splitting $\Delta_v$.  The band structures are along the high-symmetry line in the MBZ.  The dotted horizontal line denotes the Fermi energy at half-filling of the first conduction band.  We choose the parameters as $U=56$ meV and $\theta=1.28^\circ$.  The legends are the same in all figures. }
	\label{Fig4}
\end{figure*}

\section{Orbital Chern insulator}

As the strong electron-electron interactions exist in the flat bands, the four-fold degeneracy of each moir\'e band may be spontaneously broken by the interaction-induced spin-splitting $\Delta_s$ and valley-splitting $\Delta_v$.  To study the effect of interaction, we phenomenologically introduce $\Delta_s$ and $\Delta_v$ in the TDBG system,  with the Hamiltonian~\cite{J.Zhu, J.Y.Lee, F.Wu2020, M.Xie}
\begin{align}
H_{sv}=\Delta_s s_z+\Delta_v\tau_z.
\end{align}
Here $s_z$ and $\tau_z$ both denote the third Pauli matrice, but are defined in the spin and valley subspace, respectively.  In Fig.~\ref{Fig4}, with the electric potential $U=56$ meV and the twist angle $\theta=1.28^\circ$, we plot the splitted moir\'e bands and the corresponding density of states (DOS) in AB-AB stacking for a set of the splittings $(\Delta_s,\Delta_v)$.  The dotted horizontal line denotes the Fermi energy position at half-filling $n=\frac{1}{2}n_s$ of the first conduction band.  Note that $w=10.54$ meV. 

Four cases are considered.  
(i) When $\Delta_s=\Delta_v=0$, the four-fold degeneracy of the first conduction band is preserved [Fig.~\ref{Fig4}(a)], so the Fermi energy at half-filling lies in the band interior.  As $\boldsymbol K$ and $\boldsymbol K'$ valleys are connected by the TRS, they have the same DOS and thus the total DOS is four times the DOS of one flavor [Fig.~\ref{Fig4}(b)].  
(ii) When $\Delta_s=0$ and $\Delta_v=6$ meV, the bands are spin-degenerate but valley-splitted [Fig.~\ref{Fig4}(c)], so the total DOS evolves into two peaks [Fig.~\ref{Fig4}(d)].  At half-filling, we can see that a gap is opened and the system is valley polarized, with the first conduction bands in $\boldsymbol K'$ valley being completely filled while those in $\boldsymbol K$ valley being empty.  Since the TRS has been broken by $\Delta_v$, the AHE would occur in this case.
(iii) When $\Delta_s=6$ meV and $\Delta_v=0$, the spin degeneracy is broken, with the upspin bands moving upwards and downspin bands moving downwards [Fig.~\ref{Fig4}(e)].  At half-filling, a gap is also opened and the system represents a spin-polarized (SP) state.  In experiment, by evaluating the $g-$factor to be around $g\simeq2$, the observed insulating phase at half-filling was  attributed to this state, where the insulating gap is further enhanced by an in-plane magnetic field~\cite{Y.Cao2020, X.Liu, M.He}.   
(iv) When $\Delta_s=\Delta_v=6$ meV, both the spin and valley degeneracies are broken.  At half-filling, there is no gap opening and the Fermi energy also lies in the band interior [Fig.~\ref{Fig4}(g)].  Because there is an overlap of the DOS of the $\boldsymbol K$ valley, downspin band and the $\boldsymbol K'$ valley, upspin band, the total DOS exhibits three peaks [Fig.~\ref{Fig4}(h)]. 

Normally, when both the splittings are larger than the moir\'e band flatness, gaps may be opened at the odd-fillings, $n=\frac{1}{4}n_s$ or $\frac{3}{4}n_s$.  This is just the case in TBG, where the OCI phase with $C=1$ was successfully observed in the $n=\frac{3}{4}n_s$ filling gap~\cite{A.L.Sharpe, M.Serlin}.  However, in TDBG, the flatness of the first conduction band may be large and can reach $10\sim30$ meV in the region that we focus on [see Appendix A].  This may lead to the closing of the gaps at odd-fillings, as the higher bands may move into the gap by the splittings.  For example, in Fig.~\ref{Fig4}(g), the second conduction band of $\boldsymbol K'$ valley, downspin flavor moves downwards into the $n=\frac{3}{4}n_s$ gap and the first valence band of $\boldsymbol K$ valley, upspin flavor moves upwards into the $n=\frac{1}{4}n_s$ gap.  These results agree well with the experiments~\cite{Y.Cao2020, X.Liu, M.He}, in which there is no insulating state observed at $n=\frac{3}{4}n_s$ filling gap, while the insulating state at $n=\frac{1}{4}n_s$ filling gap quickly disappears at the temperature less than 3 K~\cite{Y.Cao2020}, demonstrating that the gap is very small.  According to these analysis, we suggest that the nontrivial Chern insulator phase in TDBG may only appear at half-filling $n=\frac{1}{2}n_s$ of the first conduction band when the condition $\Delta_v>\frac{1}{2}w+\Delta_s$ is satisfied, corresponding to the case of Fig.~\ref{Fig4}(c).

\begin{figure*}
	\centering
	\includegraphics[width=18cm]{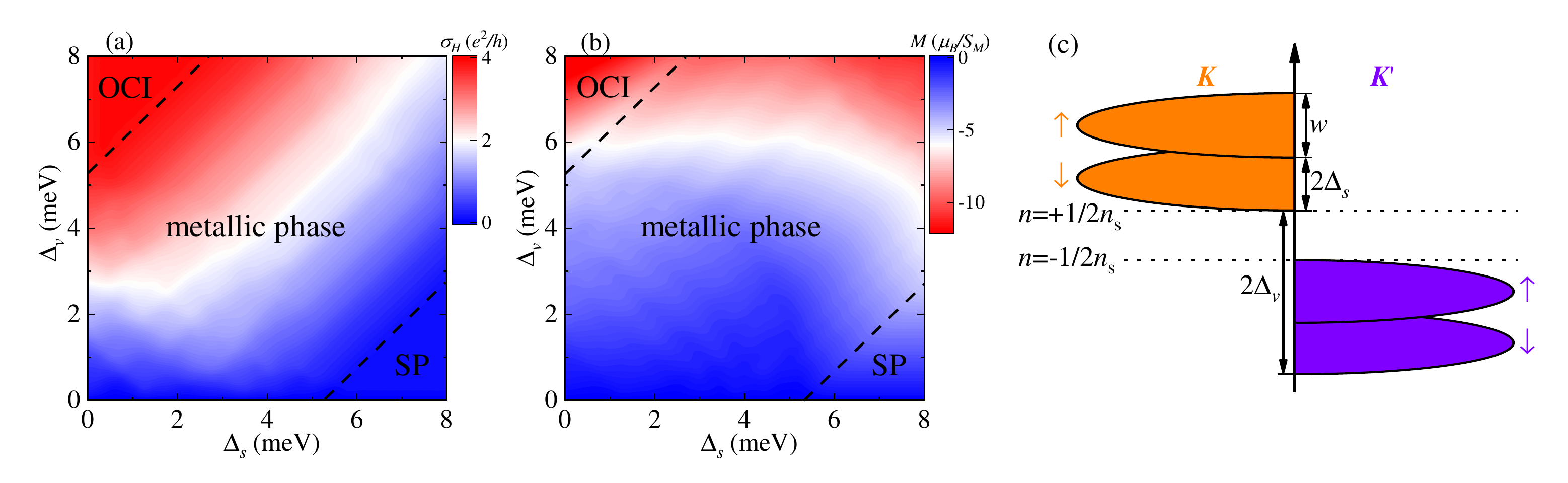}
	\caption{(Color online) The AHC $\sigma_H$ (a) and OM $M$ (b) of TDBG in the parametric space of the splittings $(\Delta_s,\Delta_v)$ with AB-AB stacking.  The dashed lines denote the phase boundaries, separating three phases: OCI phase, metallic phase and SP phase.  The Fermi energy is pinned at half-filling of the first conduction band.  If a gap is opened at the half-filling, the Fermi energy is chosen to lie at the bottom of the gap.  The parameters are taken as $U=56$ meV and $\theta=1.28^\circ$.  (c) The schematics of the valley-splitted and spin-splitted moir\'e bands.  The Fermi energy positions at $n=\pm\frac{1}{2}n_s$ filling are denoted by the dotted lines.}
	\label{Fig5}
\end{figure*}  
 
Next we study the dependence of the AHC and OM on the splittings, $\Delta_s$ and $\Delta_v$.  The AHC $\sigma_H$ is calculated by the famous TKNN formula, which expresses $\sigma_H$ as an integration of the Berry curvature over the MBZ~\cite{D.J.Thouless}, 
\begin{align}
\sigma_H=&-\frac{e^2}{\hbar}\text{Im}
\int_{\text{MBZ}}\frac{d^2\boldsymbol k}{(2\pi)^2}
\sum_{n,n'\neq n}      
\nonumber\\
&\frac{\langle u_{n\boldsymbol k}|\frac{\partial H}{\partial k_x}|u_{n'\boldsymbol k}\rangle  
\langle u_{n'\boldsymbol k}|\frac{\partial H}{\partial k_y}|u_{n\boldsymbol k}\rangle} {(\varepsilon_{n\boldsymbol k}-\varepsilon_{n'\boldsymbol k})^2} f(\varepsilon_F-\varepsilon_{n\boldsymbol k}), 
\end{align}
and the OM $M$ is calculated as~\cite{D.Xiao, T.Thonhauser, D.Ceresoli}
\begin{align}
M=&\frac{e}{\hbar}\text{Im}
\int_{\text{MBZ}}\frac{d^2\boldsymbol k}{(2\pi)^2}
\sum_{n,n'\neq n}(\varepsilon_{n\boldsymbol k}+\varepsilon_{n'\boldsymbol k}-2\varepsilon_F)
\nonumber\\
&\times\frac{\langle u_{n\boldsymbol k}|\frac{\partial H}{\partial k_x}|u_{n'\boldsymbol k}\rangle 
\langle u_{n'\boldsymbol k}|\frac{\partial H}{\partial k_y}|u_{n\boldsymbol k}\rangle} {(\varepsilon_{n\boldsymbol k}-\varepsilon_{n'\boldsymbol k})^2} 
f(\varepsilon_F-\varepsilon_{n\boldsymbol k}), 
\end{align}
where $f(\varepsilon_F-\varepsilon_{n\boldsymbol k})$ is the Fermi-Dirac distribution function and $\varepsilon_F$ is the Fermi energy.  We use $\frac{e^2}{h}$ and $\frac{\mu_B}{S_M}$ as the unit of $\sigma_H$ and $M$, respectively, with $\mu_B$ being the Bohr magneton.  The OM can be separated into two parts $M=M_1+M_2$~\cite{J.Zhu}, 
\begin{align} 
M_1=&\frac{e}{\hbar}\text{Im}
\int_{\text{MBZ}}\frac{d^2\boldsymbol k}{(2\pi)^2}
\sum_{n,n'\neq n}(\varepsilon_{n\boldsymbol k}+\varepsilon_{n'\boldsymbol k})
\nonumber\\
&\times\frac{\langle u_{n\boldsymbol k}|\frac{\partial H}{\partial k_x}|u_{n'\boldsymbol k}\rangle 
\langle u_{n'\boldsymbol k}|\frac{\partial H}{\partial k_y}|u_{n\boldsymbol k}\rangle} {(\varepsilon_{n\boldsymbol k}-\varepsilon_{n'\boldsymbol k})^2} 
f(\varepsilon_F-\varepsilon_{n\boldsymbol k}), 
\end{align}
and 
\begin{align}
M_2=&\frac{e}{\hbar}\text{Im}
\int_{\text{MBZ}}\frac{d^2\boldsymbol k}{(2\pi)^2}
\sum_{n,n'\neq n}(-2\varepsilon_F)
\nonumber\\
&\times\frac{\langle u_{n\boldsymbol k}|\frac{\partial H}{\partial k_x}|u_{n'\boldsymbol k}\rangle 
\langle u_{n'\boldsymbol k}|\frac{\partial H}{\partial k_y}|u_{n\boldsymbol k}\rangle} {(\varepsilon_{n\boldsymbol k}-\varepsilon_{n'\boldsymbol k})^2} 
f(\varepsilon_F-\varepsilon_{n\boldsymbol k}). 
\end{align}
The above equations show that when the Fermi energy $\varepsilon_F$ lies in the gap, $M_1$ is independent of $\varepsilon_F$, while $M_2$ exhibits a linear dependence on $\varepsilon_F$.  In particular, $M_2$ is closely related to the edge states as its coefficient is proportional to the Chern number in the gap, $\frac{dM_2}{d\varepsilon_F}=\frac{e}{2\pi\hbar}C_\delta$.  

We show the AHC $\sigma_H$ and OM $M$ of AB-AB stacking in Figs.~\ref{Fig5}(a) and (b), respectively.  The splittings $\Delta_s$ and $\Delta_v$ are varied from 0 to 8 meV.  The Fermi energy is pinned at half-filling of the first conduction band.  If a gap is opened at  half-filling, the Fermi energy is chosen to lie at the bottom of the gap.  We can see that when $\Delta_v=0$ and the TRS is preserved, both $\sigma_H$ and $M$ vanish, due to the opposite contributions from the $\boldsymbol K$ and $\boldsymbol K'$ valleys.  For a fixed $\Delta_s$, when $\Delta_v$ increases, more electronic states in $\boldsymbol K'$ valley than $\boldsymbol K$ valley are occupied.  Correspondingly, $\sigma_H$ increases from zero, while $M$ decreases from zero to a large negative value.

\begin{figure*}
	\centering
	\includegraphics[width=13cm]{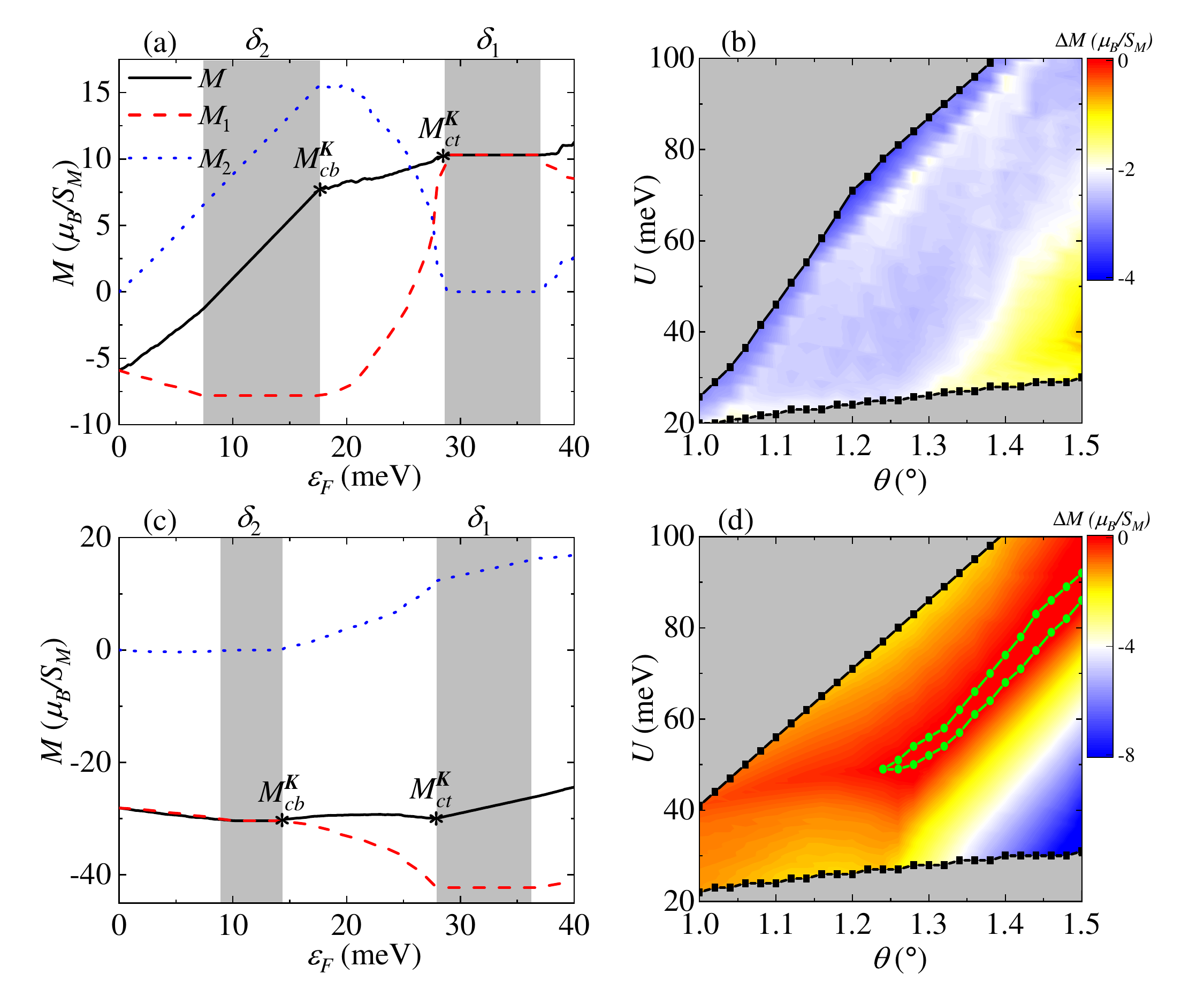}
	\caption{(Color online) The single-flavor OM $M$ in $\boldsymbol K$ valley and $\Delta M=M_{cb}^{\boldsymbol K}-M_{ct}^{\boldsymbol K}$ of TDBG with AB-AB stacking (a)-(b) and AB-BA stacking (c)-(d).  In (a) and (c), the parameters are taken as $U=56$ meV and $\theta=1.28^\circ$.  The red dashed and blue dotted lines denote $M_1$ and $M_2$, respectively.  The gray stripes indicate the energy gaps $\delta_1$ and $\delta_2$.  The extremal OMs $M_{cb}^{\boldsymbol K}$ and $M_{ct}^{\boldsymbol K}$ are marked by the asterisks.  In (b) and (d), we focus on the $(2,-2)$ phase and $(1,1)$ phase, respectively, while the gray areas are outside of the $(2,-2)$ and $(1,1)$ phase.   Both figures show that $\Delta M$ is negative, except that in a small region of (d), as highlighted by the green lines, $\Delta M$ is positive, but can reach $\sim0.1\frac{\mu_B}{S_M}$ at most.}
	\label{Fig6}
\end{figure*}

Three phases that are separated by the dashed lines can be seen in Fig.~\ref{Fig5}.  Above the phase boundary $\Delta_v=\frac{1}{2}w+\Delta_s$, 
a gap is opened at half-filling $n=\frac{1}{2}n_s$ of the first conduction band and the system enters the OCI phase.  Because both the occupied upspin and downspin band in $\boldsymbol K'$ valley have the Chern number $C_{c1}^{\boldsymbol K'}=2$ (opposite to $C_{c1}^{\boldsymbol K}=-2$), $\sigma_H$ is quantized as $4\frac{e^2}{h}$, as shown in Fig.~\ref{Fig5}(a).  But $M$ will further decrease with $\Delta_v$, due to the contributions from the edge states in the gap, as in Fig.~\ref{Fig5}(b).  As the upspin and downspin states are equally occupied in the OCI phase, the spin magnetization vanishes and therefore the total magnetization is dominated by the orbital component.  When $\Delta_v=8$ meV and $\Delta_s$ increases, $\sigma_H$ gradually deviates from the quantized value, whereas $M$ is still large and around $-10\frac{\mu_B}{S_M}$.  On the other hand, below the phase boundary $\Delta_s=\frac{1}{2}w+\Delta_v$, a gap is also opened at half-filling of the first conduction band and the system enters the SP ferromagnetic state.  As the TRS is unbroken ($\Delta_v=0$) or weakly broken ($\Delta_v\ll w$), $\sigma_H$ is zero or vanishingly small, while $M$ gives a small value, which also originates from the edge states in the gap.  Between the phase boundaries, a metallic phase is present in the parameter space, due to the finite DOS at the Fermi energy.  We note that the exact positions of the phase boundaries are dependent on the electric potential $U$ and the twist angle $\theta$, as the flatness $w$ can be effectively modulated [see Appendix A].  Although only AB-AB stacking is considered in Figs.~\ref{Fig4} and~\ref{Fig5}, similar conclusions can also be obtained for AB-BA stacking, except that the AHC would be quantized as $\sigma_H=-2\frac{e^2}{h}$ for the same $U$ and $\theta$. 

We further explore whether there is a OM reversal in the half-filling gap of TDBG when the OCI phase has been identified.  To see the behavior of the OM $M$, in Fig.~\ref{Fig6}, we plot the single-flavor $M$ in $\boldsymbol K$ valley (no splittings) as a function of the Fermi energy $\varepsilon_F$ with AB-AB stacking (a) and AB-BA stacking (c) when the parameters $U=56$ meV and $\theta=1.28^\circ$.  The red dashed and blue dotted lines denote the separated $M_1$ and $M_2$ contributions, respectively.  As $\boldsymbol K$ and $\boldsymbol K'$ valleys are TR counterparts, their OM contributions are opposite in sign.  In Fig.~\ref{Fig6}(a), we observe that $M$ in $\boldsymbol K$ valley keeps unchanged in $\delta_1$ gap as the Chern number $C_{\delta_1}^{\boldsymbol K}=0$, and increases linearly in $\delta_2$ gap as $C_{\delta_2}^{\boldsymbol K}=2$, while in Fig.~\ref{Fig6}(c), $M$ in $\boldsymbol K$ valley increases linearly in $\delta_1$ gap as $C_{\delta_1}^{\boldsymbol K}=1$ and remains unchanged in $\delta_2$ gap as $C_{\delta_2}^{\boldsymbol K}=0$.   Note that due to the absence of the electron-hole symmetry in TDBG, $M$ does not vanish at zero Fermi energy.  This is in sharp contrast with TBG, where the electron-hole symmetry is well preserved and $M$ always vanishes at zero Fermi energy~\cite{J.Zhu}.  

When the splittings are present in TDBG, the total OMs require to sum over all spin and valley flavors.  For the OCI at $n=\pm\frac{1}{2}n_s$ filling, with the Fermi energy positions being denoted by the dotted lines in Fig.~\ref{Fig5}(c), the OM in each flavor can be easily obtained.  For example, we have $M_{+\frac{1}{2}n_s}^{\boldsymbol K\uparrow}=M_{cb}^{\boldsymbol  K}-\frac{e}{2\pi\hbar}C_{\delta_2}^{\boldsymbol K} 2\Delta_s$ and $M_{-\frac{1}{2}n_s}^{\boldsymbol K\uparrow}=M_{cb}^{\boldsymbol K}-\frac{e}{2\pi\hbar}C_{\delta_2}^{\boldsymbol K} (2\Delta_v-w)$.  

For AB-AB stacking, the total OMs are 
\begin{align}
M_{+\frac{1}{2}n_s}=&2\Delta M-\frac{e}{\pi\hbar}C_{\delta_2}^{\boldsymbol K}\Delta_s,
\\
M_{-\frac{1}{2}n_s}=&2\Delta M-\frac{e}{\pi\hbar} C_{\delta_2}^{\boldsymbol K} (2\Delta_v-\Delta_s-w),  
\end{align} 
where $\Delta M=M_{cb}^{\boldsymbol K}-M_{ct}^{\boldsymbol K}$, with $M_{cb}^{\boldsymbol K}$ and $M_{ct}^{\boldsymbol K}$ denoting the extremal OM with the Fermi energy being located at the band bottom and top, respectively.  Eq.~(17) tells us that $M_{-\frac{1}{2}n_s}$ decreases with $\Delta_v$, but increases with $\Delta_s$, as observed in the top left of Fig.~\ref{Fig5}(b).  The difference between $M_{+\frac{1}{2}n_s}$ and $M_{-\frac{1}{2}n_s}$ is 
\begin{align}
M_{+\frac{1}{2}n_s}-M_{-\frac{1}{2}n_s}
=\frac{e}{\pi\hbar} C_{\delta_2}^{\boldsymbol K} (2\Delta_v-2\Delta_s-w),  
\end{align}
which is positive when taking into account the condition for the OCI, $\Delta_v>\frac{1}{2}w+\Delta_s$.  In Fig.~\ref{Fig6}(a), we can see that $0<M_{cb}^{\boldsymbol K}<M_{ct}^{\boldsymbol K}$ and thus  $M_{-\frac{1}{2}n_s}<M_{+\frac{1}{2}n_s}<0$.  This means that the OM will increase from the bottom to the top of the half-filling gap, but remains negative.  We further check this in the whole $(2,-2)$ phase.  As the total OMs depend heavily on $\Delta M$, we plot $\Delta M$ in Fig.~\ref{Fig6}(b), where $\Delta M$ is always negative, indicating that there is no OM reversal in AB-AB stacking. 

For AB-BA stacking, the total OMs are
\begin{align}
M_{+\frac{1}{2}n_s}=&2\Delta M-\frac{e}{\pi\hbar}C_{\delta_1}^{\boldsymbol K} (2\Delta_v-\Delta_s-w), 
\\
M_{-\frac{1}{2}n_s}=&2\Delta M
-\frac{e}{\pi\hbar}C_{\delta_1}^{\boldsymbol K} \Delta_s.  
\end{align} 
Their difference is 
\begin{align}
M_{+\frac{1}{2}n_s}-M_{-\frac{1}{2}n_s}
=-\frac{e}{\pi\hbar} C_{\delta_1}^{\boldsymbol K} (2\Delta_v-2\Delta_s-w),
\end{align}
which is negative for the OCI state.  Fig.~\ref{Fig6}(c) shows that $M_{cb}^{\boldsymbol K}<M_{ct}^{\boldsymbol K}<0$, so we can identify that $M_{+\frac{1}{2}n_s}<M_{-\frac{1}{2}n_s}<0$, meaning that the OM will decrease from the bottom to the top of the half-filling gap, but again remains negative.  In Fig.~\ref{Fig6}(d), we also check $\Delta M$ in the whole $(1,1)$ phase.  It shows that $\Delta M$ is mostly negative, except for a small region where $\Delta M$ becomes positive, as has been highlighted by the green lines.  However, in this region, $\Delta M$ can only reach $\sim0.1\frac{\mu_B}{S_M}$ at most.  Considering that $\frac{e\cdot\text{meV}}{\pi\hbar}=0.884\frac{\mu_B}{S_M}$ and $2\Delta_v-\Delta_s-w>\Delta_s$, a sufficiently weak $\Delta_s\sim0.23$ meV can make $M_{\pm\frac{1}{2}n_s}$ remain negative.  Thus we suggest that the OMs are also negative and there is no OM reversal in AB-BA stacking.  

In Ref.~\cite{J.Zhu}, by studying the magnetic property in TBG, the authors expected that the sign reversal of the OM is common in the large gap OCI.  Here we have demonstrated that this conclusion does not hold in the OCI state of TDBG, which may be attributed to the specific band topologies in TDBG.  So the necessary conditions for the OM reversal in the OCI state based on the moir\'e flat-band systems need more investigations.

\section{Discussions and Summaries}

When comparing with the experiments~\cite{Y.Cao2020, X.Liu, M.He}, we find that the typical electric potential performed on TDBG is higher than that used in our theoretical calculations.  This may be attributed to the fact that in our model, the uniform electric potential drop is assumed between neighboring layers, but in real samples, the uniform electric potential drop cannot exist, because the separation between the double bilayer graphene is evidently larger than the separation between the two layers of one bilayer graphene.  Nevertheless, the effect of the electric potential in TDBG can still be qualitatively captured by the theoretical model.  

We make some comparisons of the topological moir\'e bands between TBG and TDBG.  In TBG, the observation of the flat moir\'e bands needs to fix the twist angle to the specific magic angle, $\theta\sim1.1^\circ$.  The nontrivial band topology requires to perfectly align the TBG system with the hexagonal boron nitride cladding layers~\cite{M.Serlin, J.Jung2015, N.Bultinck}, as to break the $C_{2z}$ symmetry between the two sublattices and acquire a finite mass for the Dirac cone.  These conditions are rather strict constraints in experiment.  Here in TDBG, the flat bands can exist in a large twist-angle range, as is shown in the phase diagrams and has been demonstrated in experiment~\cite{Y.Cao2020, X.Liu, M.He}.  Because both the twist angle and electric potential can be controlled in experiment, this makes the band Chern number in TDBG be effectively modulated. 

To summary, in this paper, we have investigated the phase diagram and OCI in TDBG modulated by the twist angle and the electric field.  As the stacking type plays an important role in determining the band topology of TDBG, we find that it can be inferred by judging the valley Chern number with the reversed direction of the electric potential.  The appearance of the OCI in TDBG requires the strong valley splitting to open a gap at half-filling of the first conduction band.  The experiments~\cite{Y.Cao2020, X.Liu, M.He} and the Hatree-Fock calculations~\cite{J.Y.Lee} pointed to the SP state at half-filling, due to the strong spin splittings by the correlation effect.  Therefore the realization of the strong valley splitting in TDBG may require more delicate conditions, which need more theoretical and experimental studies in the future. 

\begin{figure*}
	\centering
	\includegraphics[width=18cm]{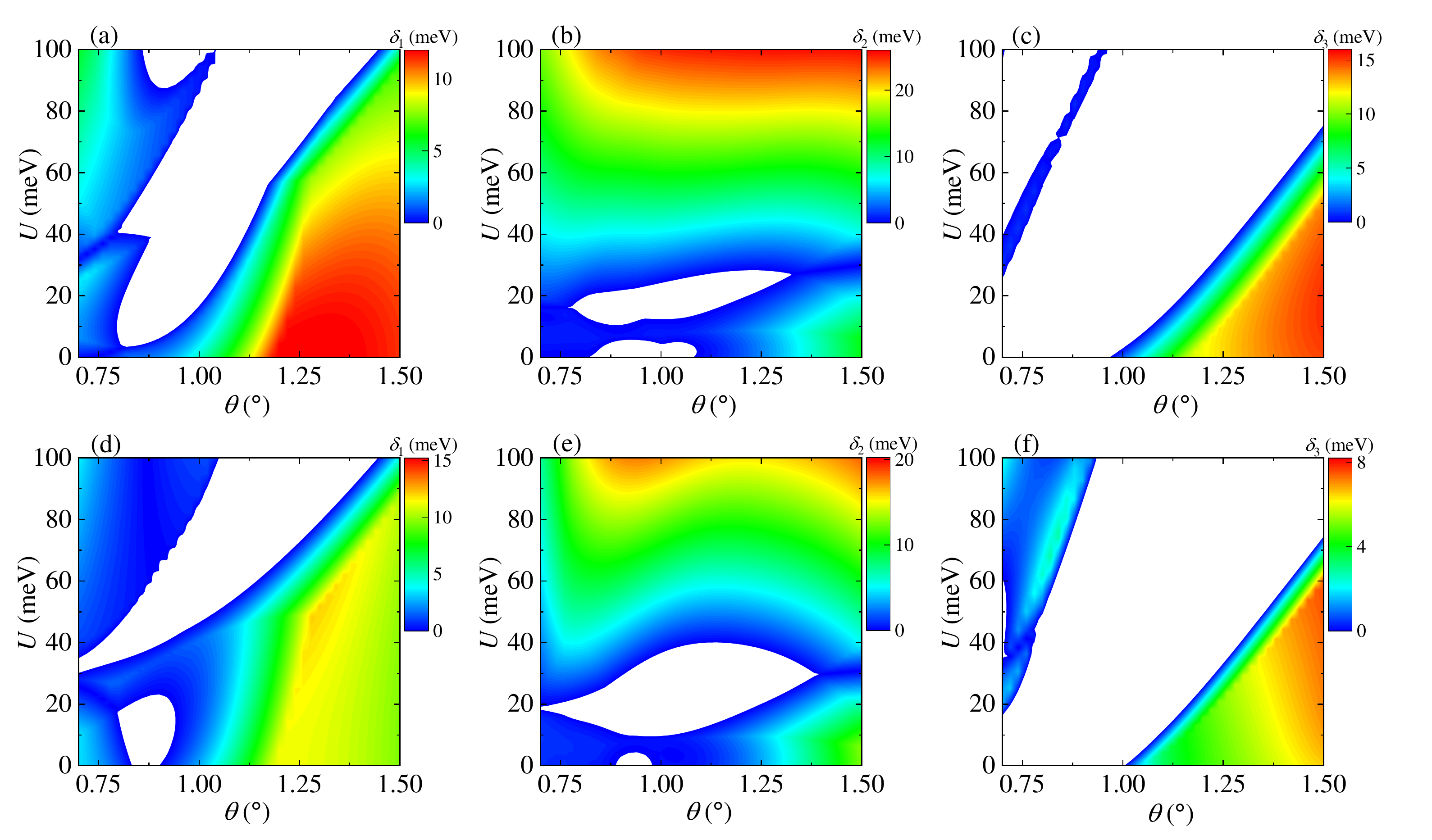}
	\caption{(Color online) The bandgaps $\delta_1$, $\delta_2$ and $\delta_3$ of TDBG in the parametric space $(\theta,U)$ with AB-AB stacking (a)-(c) and AB-BA stacking (d)-(f). $\delta_1$ is the bandgap between the first conduction and second conduction band, $\delta_2$ is the bandgap between the first conduction and valence band, and $\delta_3$ is the bandgap between the first valence and second valence band. }
	\label{FigA7}
\end{figure*} 

\begin{figure*}
	\centering
	\includegraphics[width=12.5cm]{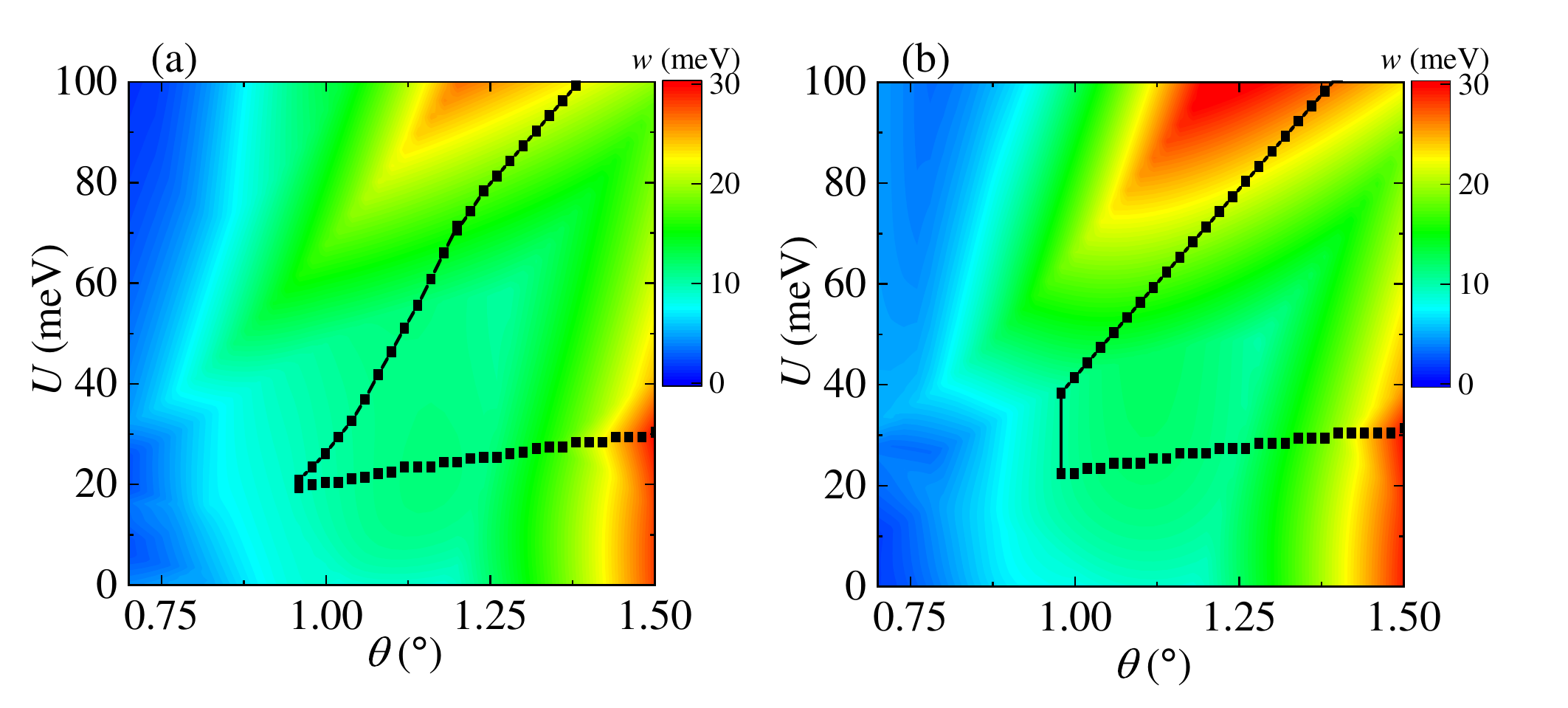}
	\caption{(Color online) The flatness $w$ of the first conduction band of TDBG in the parametric space $(\theta,U)$ with AB-AB stacking (a) and AB-BA stacking (b).  The regions that we focus on the OCI are inside of the black lines in each figure. }
	\label{FigA8}
\end{figure*}

\section{Acknowledgments}

This work was supported by NSFC (Grants No. 11704157, No. 11804122 and No. 11905054), and the Fundamental Research Funds for the Central Universities of China.

\section{Appendix}

\subsection{Bandgaps and Flatness}

We calculate the bandgaps $\delta_1$, $\delta_2$ and $\delta_3$ of TDBG, which are defined as
\begin{align}
&\delta_1=\text{min}(\varepsilon_{2c})-\text{max}(\varepsilon_{1c}),
\\
&\delta_2=\text{min}(\varepsilon_{1c})-\text{max}(\varepsilon_{1v}),
\\
&\delta_3=\text{min}(\varepsilon_{1v})-\text{max}(\varepsilon_{2v}),
\end{align}
and the flatness of the first conduction band
\begin{align}
w=\text{max}(\varepsilon_{1c})-\text{min}(\varepsilon_{1c}).
\end{align}

In Fig.~\ref{FigA7}, the contour plots of the bandgaps are presented.  When the neighboring bands are overlapped, there is no direct bandgap opening and the bandgap becomes negative.  We can see that the regions for the overlapped bands are roughly the same for AB-AB stacking and AB-BA stacking.  Note that for AB-AB stacking, the obtained bandgap results are consistent with those in Figs.~7(a)-(c) of Ref.~\cite{J.Y.Lee}.  For $\delta_2$, the regions appear in the bottom with low $U$ [Figs.~\ref{FigA7}(b) and (e)], while for $\delta_1$ and $\delta_3$, the overlapped bands can span quite a large region in the parametric space [Figs.~\ref{FigA7}(a), (c), (d) and (f)].  Thus we arrive at the conclusion that the overlapped bands related to the first conduction and first valence bands are quite common for the small twist-angle TDBG modulated by the electric potential. 

\begin{figure*}
	\centering
	\includegraphics[width=13.4cm]{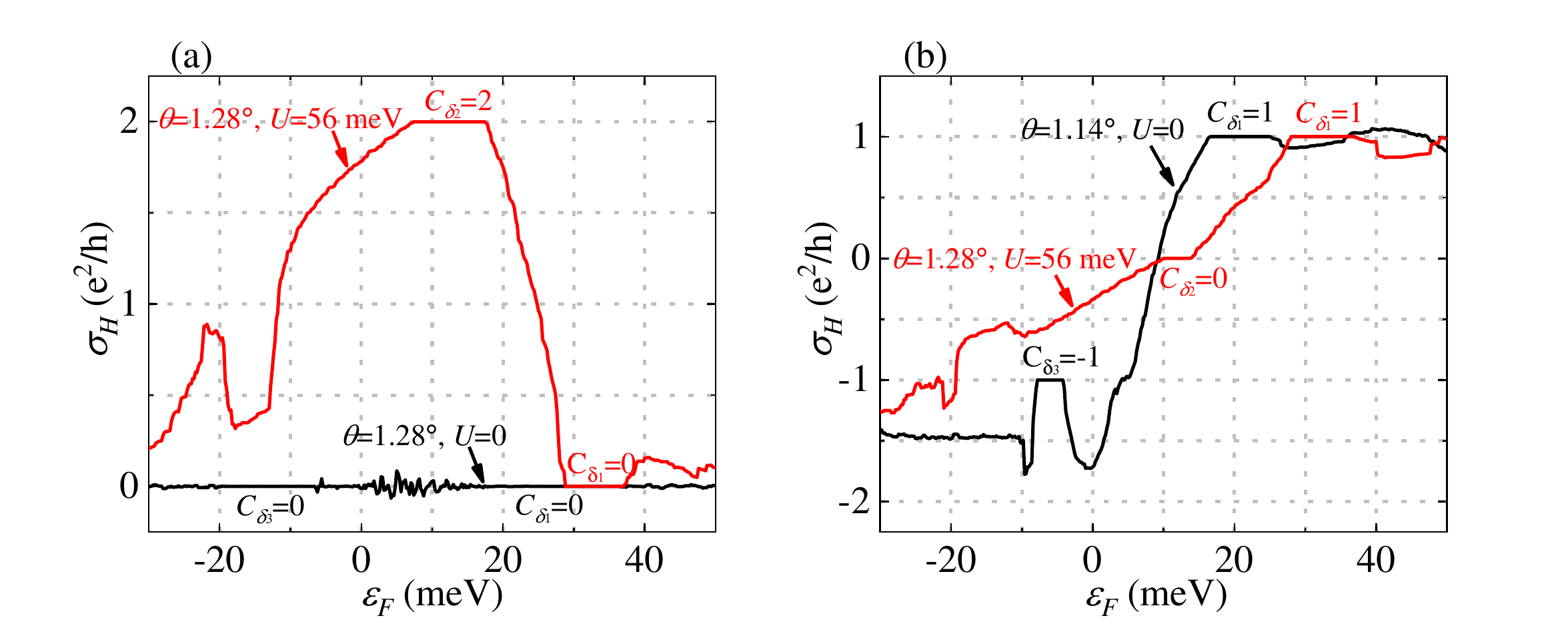}
	\caption{(Color online) The AHC $\sigma_H$ (in unit of $\frac{e^2}{h}$) calculated by using the TKNN formula [Eq.~(12)], with the parameters in (a) and (b) being the same as those in Figs.~\ref{Fig2}(c)-(d) and Figs.~\ref{Fig2}(g)-(h), respectively.  When the neighboring bands are overlapped, $\sigma_H$ will not be quantized, as in both (a) and (b), there is no $C_{\delta_3}$ plateau of the red line and no $C_{\delta_2}$ plateau of the black line.}
	\label{FigA9}
\end{figure*}

In Fig.~\ref{FigA8}, the contour plots of the flatness are presented.  The regions that we focus on the OCI, the $(2,-2)$ phase in Fig.~\ref{FigA8}(a) and (1,1) phase in Fig.~\ref{FigA8}(b), are inside the black lines.  We can see that in these regions, the flatness $w$ can reach $10\sim30$ meV, ensuring that the bandgap cannot be opened at the odd fillings, $n=\frac{1}{4}n_s$ or $n=\frac{3}{4}n_s$, by the splittings.

\subsection{Chern number determination from the TKNN formula}

By using the TKNN formula [Eq.~(12)], the AHC $\sigma_H$ of the system can be calculated as a function of the Fermi energy $\varepsilon_F$.  In Fig.~\ref{FigA9}, we plot the calculated $\sigma_H$, with the parameters in (a) and (b) chosen the same as those in Figs.~\ref{Fig2}(c)-(d) and (g)-(h), respectively. 

Comparing the Chern number determination from the TKNN formula and Fukui's algorithm, we can see that when the direct gap dominates the system, the results are the same.  But when the neighboring bands are overlapped, the results are different, as $\sigma_H$ will not be quantized in the TKNN formula.  This is clearly seen in both Figs.~\ref{FigA9}(a) and (b), where there is no  $C_{\delta_3}$ plateau of the red line and no $C_{\delta_2}$ plateau of the black line. 

For example, in Fig.~\ref{FigA9}(a) of the red line, with $\theta=1.28^\circ$ and $U=56$ meV, we observe that $C_{\delta_1}=0$ and $C_{\delta_2}=2$.  Then the Chern number of the first conduction band is determined as $C_{c1}=C_{\delta_1}-C_{\delta_2}=-2$.  This Chern number value is consistent with that obtained from the Fukui's algorithm, as labeled in Fig.~\ref{Fig2}(d).  On the other hand, when the negative $\delta_3-$gap is present that the first valence and second valence bands are overlapped, which is evidently seen in the moir\'e band structure in Fig.~\ref{Fig2}(d), no quantized $\sigma_H$ is observed.  Consequently, the Chern number of the first valence band, $C_{v1}$, is not well judged from the TKNN formula.  Similar cases can also be seen in other lines of Fig.~\ref{FigA9}.  However, even with the presence of the overlapped bands, we suggest that the Chern number can still be well determined from the Fukui's algorithm, as long as the neighboring moir\'e bands do not touch with each other.


\begin{references}
	
\bibitem{Y.Cao2018a}
Y. Cao, V. Fatemi, A. Demir, S. Fang, S. L. Tomarken, J. Y. Luo, J. D. Sanchez-Yamagishi, K. Watanabe, T. Taniguchi, E. Kaxiras, R. C. Ashoori, and P. J. Herrero,   
Nature (London) {\bf556}, 80 (2018).

\bibitem{Y.Cao2018b}
Y. Cao, V. Fatemi, S. Fang, K. Watanabe, T. Taniguchi, E. Kaxiras, and P. J. Herrero, 
Nature (London) {\bf556}, 43 (2018).

\bibitem{M.Yankowitz}
M. Yankowitz, S. Chen, H. Polshyn, Y. Zhang, K. Watanabe, T. Taniguchi, D. Graf, A. F. Young, and C. R. Dean, 
Science {\bf363}, 1059 (2019).	

\bibitem{A.L.Sharpe}
A. L. Sharpe, E. J. Fox, A. W. Barnard, J. Finney, K. Watanabe, T. Taniguchi, M. A. Kastner, and D. G. Gordon, 
Science {\bf365}, 608 (2019). 

\bibitem{M.Serlin}
M. Serlin, C. L. Tschirhart, H. Polshyn, Y. Zhang, J. Zhu, K. Watanabe, T. Taniguchi, L. Balents, and A. F. Young, 
Science {\bf367}, 900 (2020). 

\bibitem{R.Bistritzer} 
R. Bistritzer and A. MacDonald, 
Proc. Natl. Acad. Sci. USA {\bf108}, 12233 (2011). 	

\bibitem{B.Roy}
B. Roy and V. Juricic,
Phys. Rev. B {\bf99}, 121407(R) (2019). 

\bibitem{B.Lian}
B. Lian, Z. Wang, and B. A. Bernevig, 
Phys. Rev. Lett. {\bf122}, 257002 (2019). 

\bibitem{F.Wu2018}
F. Wu, A. H. MacDonald, and I. Martin, 
Phys. Rev. Lett. {\bf121}, 257001 (2018). 

\bibitem{Y.H.Zhang}
Y. H. Zhang, D. Mao, Y. Cao, P. Jarillo-Herrero, and T. Senthil, 
Phys. Rev. B {\bf99}, 075127 (2019). 

\bibitem{G.W.Burg}
G. W. Burg, J. Zhu, T. Taniguchi, K. Watanabe, A. H. MacDonald, and E. Tutuc, 
Phys. Rev. Lett. {\bf123}, 197702 (2019).

\bibitem{C.Shen}
C. Shen, Y. Chu, Q. Wu, N. Li, S. Wang, Y. Zhao, J. Tang, J. Liu, J. Tian, K. Watanabe, T. Taniguchi, R. Yang, Z. Meng, D. Shi, O. V. Yazyev, and G. Zhang
Nat. Phys. {\bf16}, 520 (2020). 

\bibitem{Y.Cao2020}
Y. Cao, D. R. Legrain, O. R. Bigorda, J. M. Park, K. Watanabe, T. Taniguchi, and P. J. Herrero, 
Nature (London) {\bf583}, 215 (2020).  

\bibitem{X.Liu}
X. Liu, Z. Hao, E. Khalaf, J. Y. Lee, Y. Ronen, H. Yoo, D. H. Najafabadi, K. Watanabe, T. Taniguchi, A. Vishwanath, and P. Kim, 
Nature (London) {\bf583}, 221 (2020).  

\bibitem{M.He}
M. He, Y. Li, J. Cai, Y. Liu, K. Watanabe, T. Taniguchi, X. Xu, and M. Yankowitz, 
arXiv: 2002.08904.

\bibitem{G.Chen2019a}
G. Chen, L. Jiang, S. Wu, B. Lyu, H. Li, B. L. Chittari, K. Watanabe, T. Taniguchi, Z. Shi, J. Jung, Y. Zhang, and F. Wang, 
Nat. Phys. {\bf15}, 237 (2019).

\bibitem{G.Chen2019b}
G. Chen, A. L. Sharpe, P. Gallagher, I. T. Rosen, E. J. Fox, L. Jiang, B. Lyu, H. Li, K. Watanabe, T. Taniguchi, J. Jung, Z. Shi, D. Goldhaber-Gordon, Y. Zhang, and F. Wang, 
Nature (London) {\bf572}, 215 (2019).

\bibitem{M.Koshino2019}
M. Koshino, 
Phys. Rev. B {\bf99}, 235406 (2019). 

\bibitem{I.Martin}
I. Martin, Y. M. Blanter, and A. F. Morpurgo, 
Phys. Rev. Lett. {\bf100}, 036804 (2008).

\bibitem{F.Zhang}
F. Zhang, A. H. MacDonald, and E. J. Mele, 
Proc. Natl. Acad. Sci. USA {\bf110}, 10546 (2013).

\bibitem{C.Z.Chang}
C. Z. Chang, J. Zhang, X. Feng, J. Shen, Z. Zhang, M. Guo, K. Li, Y. Ou, P. Wei, L. L. Wang, Z.-Q. Ji, Y. Feng, S. Ji, X. Chen, J. Jia, X. Dai, Z. Fang, S.-C. Zhang, K. He, Y. Wang, L. Lu, X.C. Ma and Q. K Xue, 
Science {\bf340}, 167 (2013).

\bibitem{Haldane}
F. D. M. Haldane, 
Phys. Rev. Lett. {\bf61}, 2015 (1988).

\bibitem{J.Zhu}
J. Zhu, J. J. Su, A. H. MacDonald, 
Phys. Rev. Lett. {\bf125}, 227702 (2020). 

\bibitem{H.Polshyn}
H. Polshyn, J. Zhu, M. A. Kumar, Y. Zhang, F. Yang, C. L. Tschirhart, M. Serlin, K. Watanabe, T. Taniguchi, A. H. MacDonald, and A. F. Young, arxiv: 2004.11353. 

\bibitem{Neto}
A. H. Castro Neto, F. Guinea, N. M. R. Peres, K. S. Novoselov,
and A. K. Geim, 
Rev. Mod. Phys. {\bf81}, 109 (2009).

\bibitem{M.Koshino2018}
M. Koshino, N. F. Q. Yuan, T. Koretsune, M. Ochi, K. Kuroki, and L. Fu, 
Phys. Rev. X {\bf8}, 031087 (2018). 	
	
\bibitem{J.Y.Lee}
J. Y. Lee, E. Khalaf, S. Liu, X. Liu, Z. Hao, P. Kim, and A.
Vishwanath, 
Nat. Commun. {\bf10}, 5333 (2019).
  	
\bibitem{J.Liu} 
J. Liu, Z. Ma, J. Gao, and X. Dai, 
Phys. Rev. X {\bf9}, 031021 (2019). 

\bibitem{N.R.Chebrolu}
N. R. Chebrolu, B. L. Chittari, and J. Jung, 
Phys. Rev. B {\bf99}, 235417 (2019). 

\bibitem{E.McCann}
E. McCann and M. Koshino, 
Rep. Prog. Phys. {\bf76}, 056503 (2013). 

\bibitem{J.Jung2014}
J. Jung and A. H. MacDonald, 
Phys. Rev. B {\bf89}, 035405 (2014). 

\bibitem{D.Xiao}
D. Xiao, J. Shi, and Q. Niu, 
Phys. Rev. Lett. {\bf95}, 137204 (2005).

\bibitem{T.Fukui}
T. Fukui, Y. Hatsugai, and H. Suzuki, 
J. Phys. Soc. Jpn {\bf74}, 1674 (2005). 

\bibitem{Y.X.Wang}
Y. X. Wang, F. Li, and Y. M. Wu, 
EPL {\bf105}, 17002 (2014). 

\bibitem{F.Wu2020}
F. Wu and S. Das Sarma, 
Phys. Rev. B {\bf101}, 155149 (2020). 

\bibitem{M.Xie}
M. Xie and A. H. MacDonald, 
Phys. Rev. Lett. {\bf124}, 097601 (2020). 

\bibitem{D.J.Thouless}
D. J. Thouless, M. Kohmoto, M. P. Nightingale, and M. den Nijs, 
Phys. Rev. Lett. {\bf49}, 405 (1982). 

\bibitem{T.Thonhauser}
T. Thonhauser, D. Ceresoli, D. Vanderbilt, and R. Resta,
Phys. Rev. Lett. {\bf95}, 137205 (2005).

\bibitem{D.Ceresoli}
D. Ceresoli, T. Thonhauser, D. Vanderbilt, and R. Resta,
Phys. Rev. B {\bf74}, 024408 (2006).  

\bibitem{J.Jung2015}
J. Jung, A. M. Dasilva, A. H. MacDonald, and S. Adam,   
Nat. Comm. {\bf6}, 6308 (2015).

\bibitem{N.Bultinck}
N. Bultinck, S. Chatterjee, and M. P. Zaletel, 
Phys. Rev. Lett. {\bf124}, 166601 (2020). 

\end{references}
\end{document}